\documentclass[aps,prl,reprint,superscriptaddress]{revtex4-1}
\usepackage{graphicx}
\usepackage{color}

\begin{document}


\title{Local connectivity modulates multi-scale relaxation dynamics
in a metallic glass-forming system}

\author{Z. W. Wu}
\email[]{zwwu@pku.edu.cn}
\affiliation{International Center for Quantum Materials,
School of Physics, Peking University, Beijing 100871, China}
\affiliation{Collaborative Innovation Center of Quantum Matter, Beijing, China}
\author{W. H. Wang}
\affiliation{Institute of Physics, Chinese Academy of Sciences,
Beijing 100190, China}
\author{Limei Xu}
\email[]{limei.xu@pku.edu.cn}
\affiliation{International Center for Quantum Materials,
School of Physics, Peking University, Beijing 100871, China}
\affiliation{Collaborative Innovation Center of Quantum Matter, Beijing, China}


\date{\today}

\begin{abstract}
The structural description for the intriguing link between
the fast vibrational dynamics and slow diffusive dynamics
in glass-forming systems is one of the most challenging issues
in physical science. Here, in a model of metallic supercooled liquid,
we find that local connectivity as an atomic-level structural order
parameter tunes the short-time vibrational excitations of the
icosahedrally coordinated particles and meanwhile modulates their
long-time relaxation dynamics changing from stretched to compressed
exponentials, denoting a dynamic transition from subdiffusive to
hyperdiffusive motions of such particles. Our result indicates that
long-time dynamics has an atomic-level structural origin which is
related to the short-time dynamics, thus suggests a structural
bridge to link the fast vibrational dynamics and the slow structural
relaxation in glassy materials.
\end{abstract}


\maketitle



Supercooled liquid transforms to non-equilibrium glass accompanying
by drastically slowing down of dynamics but no obvious change in
structures, which makes the description of structure-dynamics relationships
in supercooled liquids and glasses one of the most challenging problems
in condensed matter physics~\cite{Debenedetti-nat}. A variety of liquids
under supercooling are known to exhibit unusual slow dynamics,
which is often manifested as a stretched exponential decay in the
time correlation functions~\cite{Ediger-jpc,Kob-pre,Kob-jpcm}.
This reflects a wide distribution of relaxation times typically attributed
to the existence of dynamic heterogeneities. However, in out-of-equilibrium
materials (e.g., structural glasses~\cite{Ruta-nc,Ruta-prl}, colloidal
suspensions and gels~\cite{Cip-prl,Bal-np}, and for nanoparticle
within glass former matrix~\cite{Car-prl,Guo-prl}), compressed correlation
functions decaying faster than exponential have instead been reported.
This is in sharp contrast to the diffusive or subdiffusive dynamics which
relaxes slower-than-exponential typically observed in a supercooled liquid.
It was proposed that the faster-than-exponential relaxation was due to
the release of internal stresses~\cite{Cip-prl,Bal-np} or a cooperative
motion induced by the near-vitreous solvent~\cite{Car-prl,Guo-prl}.
However, a detailed description of the microscopic origin of the dynamics
is still lacking. A particle-level structural origin for the subdiffusive to
hyperdiffusive dynamics crossover is highly desirable.

Another striking feature of the time correlation functions is the presence
of a small valley between the short time (`microscopic') process and the
caging plateau that precedes the long time $\alpha$-relaxation. It is
suggested~\cite{An-sci,Hor-pre,Kob-prl,Hor-epj,Sas-nm} that the damping
oscillation (or valley) in the time correlation function is a time-domain
manifestation of the much-discussed boson peaks, which appears as
the inelastic process in frequency domain detected by light, x-ray, and
neutron scattering~\cite{Set-sci,Sok-prl}. Recent studies demonstrated
the existence of a direct link between the slow structural relaxation and
fast boson peak dynamics~\cite{Shin-nm,Wang-prl}, which correlates the
diffusive inter-basin dynamics to the vibrational intra-basin dynamics in
glass-forming systems~\cite{Sco-sci}. This raises an interesting question
of whether there exists a connection between the long-time compressed
exponential decay and the short-time valley in the time correlation functions,
which is important for the fundamental understanding of the dynamics of
glassy materials. Therefore, a comprehensive study of the structural
dependence of this valley and its relationship to the faster-than-exponential
relaxations is of crucial importance to reveal the structural link between
slow structural relaxation and fast vibrational dynamics in
glass-forming systems.

In this letter, by using a model of fragile Cu$_{50}$Zr$_{50}$ metallic liquid,
we systematically investigated the dynamics of icosahedrally coordinated
particles by measuring the incoherent intermediate scattering functions.
At short-time scale, a valley is observed in the correlation function between
the ballistic regime and the caging plateau, meaning that these particles
exhibit the dynamical feature usually belonging to strong glass formers.
Further analysis unveils that the valley position and its damping amplitude
depend on the local connectivity of the studied particles. Meanwhile, as
the local connectivity increases, the long-time behavior of the intermediate
scattering function changes from stretched exponential decay to compressed one,
indicative of a dynamic crossover from subdiffusive to hyperdiffusive motions.
Our result indicates that the local connectivity bridges multi-time scales
dynamic processes in supercooled liquids and glasses, suggesting a possible
structural origin to reveal the well-recognized correlation between the fast
vibrational dynamics and the slow structural relaxation in glass-forming systems.
%


The studied Cu$_{50}$Zr$_{50}$ system consists of $N=$ 10000 atoms
with 5000 Cu and 5000 Zr atoms initially randomly distributed in a cubic box
with periodic boundary conditions. The embedded-atom method (EAM) potential
is used to characterize the interactions between atoms~\cite{Mendelev-jap}.
In our study, molecular dynamics (MD) simulations are performed using the
LAMMPS package~\cite{Plimpton-jcp} with integration time of 1 fs. Each sample
is firstly equilibrated for long time (2 ns) at temperature $T=$ 2000 K and then
cooled down to 1000 K at rate of 1 K/ps along constant pressure $P=$ 0 bar.
The Nose-Hoover thermostat and barostat is employed for the constant pressure
ensemble (NPT-ensemble). All data are collected after another long-time
equilibration of 1 ns at 1000 K.

\begin{figure}[ht]
\centering
\includegraphics[width=1.0\linewidth]{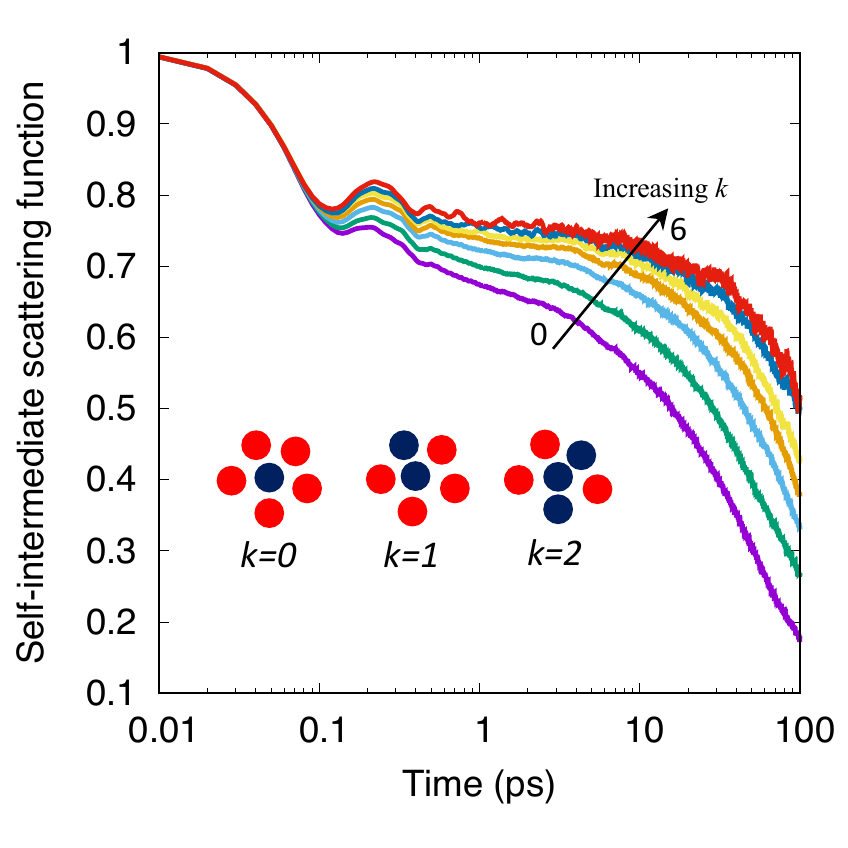}
\caption{The incoherent intermediate scattering function
(ISF) of icosahedrally coordinated particles with different
local connectivity $k$. The inset is an illustration of the local
connectivity of a particle, which is defined as the number
of its nearest-neighbors having the same local symmetry
(represented by the same color in this figure) with it.\label{allisf}}
\end{figure}

We employed the Voronoi tessellation method to characterize the local atomic
strutures in supercooled liquid~\cite{Finney-nat}. The particle with icosahedral
coordination is identified by its corresponding Voronoi index
$\langle0,0,12,0\rangle$~\cite{Sheng-nat}. The connectivity $k$ of an
icosahedrally coordinated particle is defined as the number of other icosahedrally
coordinated particles directly `connected' to it (being nearest-neighbors of each other,
as shown on the inset of Fig.~\ref{allisf})~\cite{Wu-prb,Wu-sr}.
For each $k$, we measured the corresponding incoherent intermediate scattering functions (ISF),
$F_s(q,t)=(1/N_k)\langle\sum_j\exp\{i\vec{q}\cdot[\vec{r}_j(t)-\vec{r}_j(0)]\}\rangle$,
where the summation runs over all particles with the same $k$, and
$\vec{r}_j(t)$ is the location of particle $j$ at time $t$. The value of $\vec{q}$ is chosen
as 2.8 \r{A}$^{-1}$, approximately equal to the value of the first peak position in
the structure factor, and $\langle\cdot\rangle$ denotes the ensemble average.
%


\begin{figure}[ht]
\centering
\includegraphics[width=1.0\linewidth]{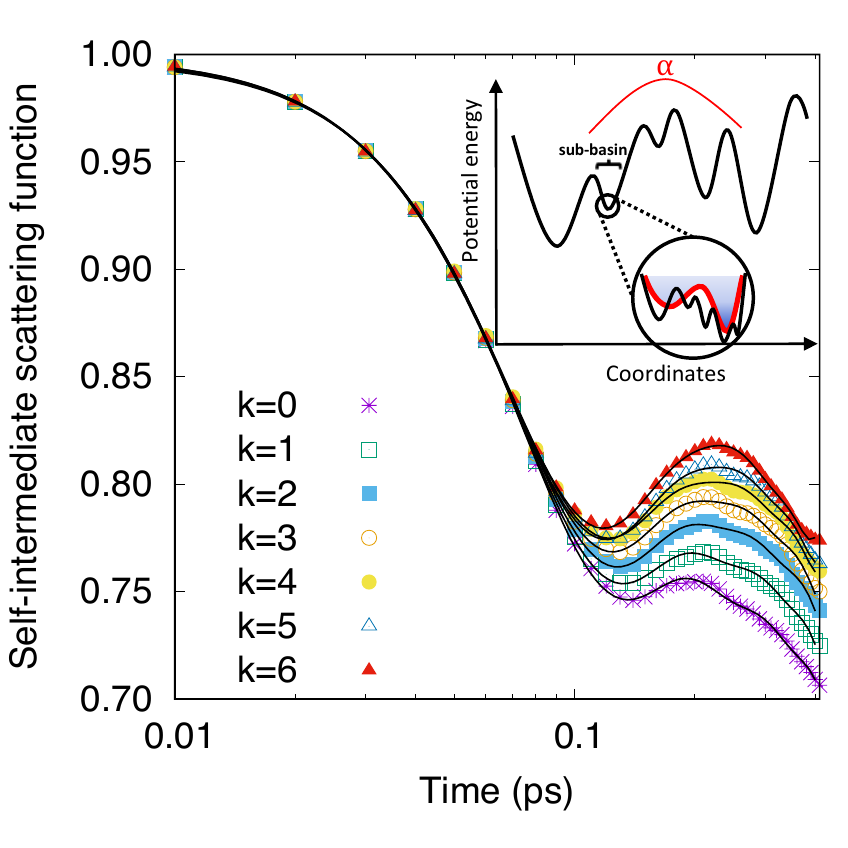}
\caption{The intermediate scattering function of particles
with different local connectivity $k$ at short-time scale.
Oscillating damping is observed in all correlation functions,
where the black solid lines are the fits of our model (Eq.~\ref{lcho}).
The inset is an illustration of the two-level-like approximation
in our model (Eq.~\ref{lcho}). The model describes the short-time
dynamics of particles rattling around their initial positions,
which correspond to vibrational excitations within a sub-basin
in the perspective of the potential energy landscape.\label{isf}}
\end{figure}

Figure~\ref{allisf} shows time dependence of the ISFs of the icosahedrally
coordinated particles with different local connectivity $k$. At short-time
scale ($t<$ 1 ps), the ISF for each $k$ exhibits a valley, a typical feature
of strong liquids~\cite{An-sci,Hor-pre,Kob-prl,Hor-epj,Sas-nm}. The amplitude
and the occurring time of the valley depend on the connectivity $k$. As $k$ increases,
the valley position shifts to shorter time accompanying with stronger amplitudes.
Meanwhile, the long-time behavior of the correlation functions displays a change
in shape from stretched exponential decay to compressed one which will be
discussed in the next sections. Such change in the long-time behavior of the
ISFs corresponds to a dynamic crossover as $k$ increases. Therefore, the
local connectivity is an atomic-level structural order parameter that can
modulate both the short-time vibrational dynamics and the long-time
relaxation dynamics of the studied particles.

At short times, 0.01 ps $\le t\le$ 0.4 ps, the ISF for different $k$ can be well
fitted by a linear combination of harmonic oscillators model (LCHO),
\begin{small}
\begin{equation}\label{lcho}
F_s(q,t)\propto \sum_{i=L,H} C_i\exp\{iq[Acos(\omega_i t+\delta_i)-Acos(\delta_i)]\},
\end{equation}
\end{small}
where $C_{L/H}$ is the weight of the harmonic oscillator with vibrational
frequency $\omega_{L/H}$ and initial phase $\delta_{L/H}$ with $L$ and $H$
denoting the low and high frequency modes respectively, and $A$ is
a fitting parameter to balance the dimension of the wave vector $q$.
The LCHO model describes the short-time dynamics of particles
rattling around their initial positions, which correspond to vibrational
excitations within a sub-basin in the perspective of the potential
energy landscape shown in the inset of Fig.~\ref{isf}. In general,
the sub-basin is rough~\cite{Char-nc} and aggregated by hierarchy
of local minima with different barriers (inset in Fig.~\ref{isf}). The
two-level-like (red filled curve in the inset of Fig.~\ref{isf}) LCHO
model is a reasonable first-step approximation for excitations
within sub-basin.

\begin{figure}[ht]
\centering
\includegraphics[width=1.0\linewidth]{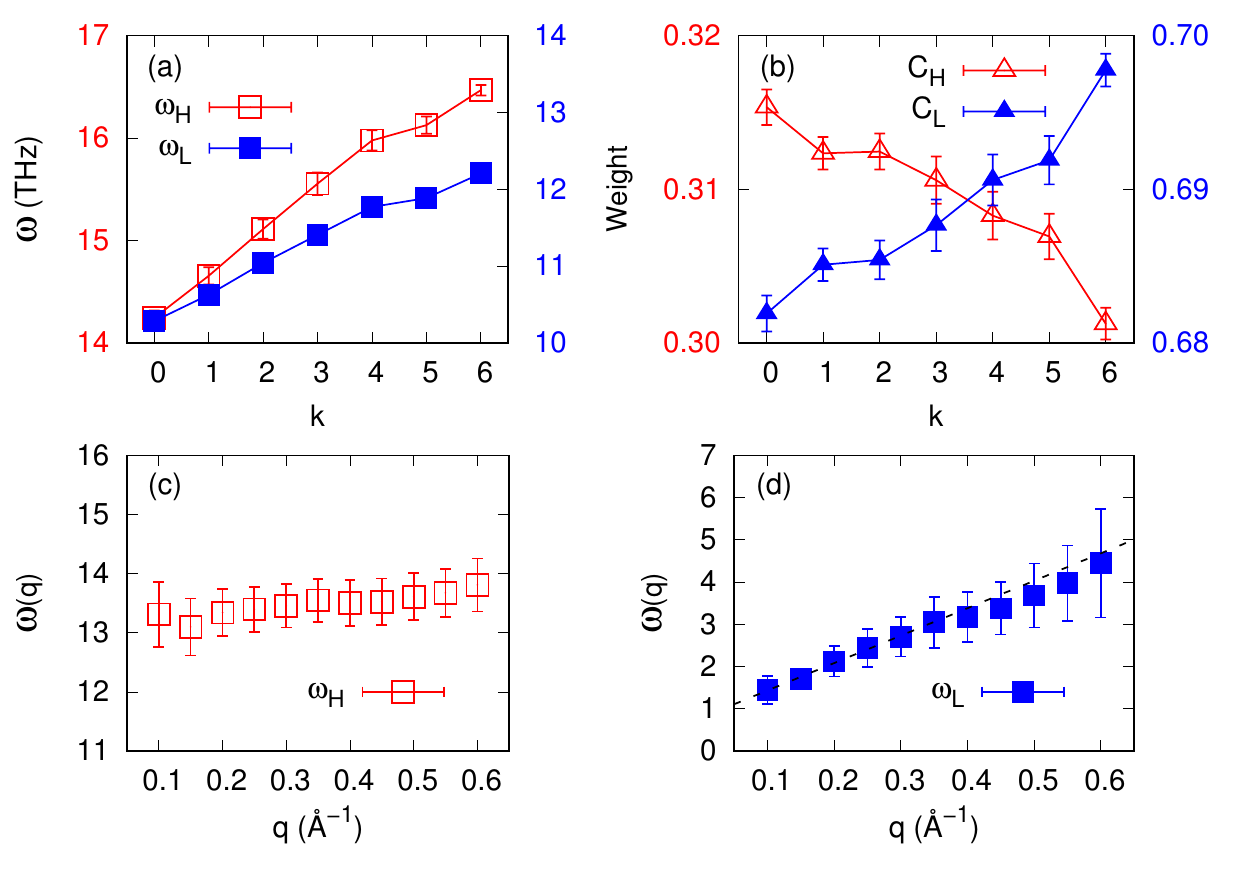}
\caption{The dependence of dynamic properties on connectivity $k$ and dispersion
relations. Both the high and low frequency modes, $\omega_H$ and $\omega_L$,
increases with increasing $k$ (a), however, the fraction of motion $C_{H/L}$
increases for $\omega_L$ and decreases for $\omega_H$ (b). The high frequency
mode $\omega_H(q)$ is approximately $q$-independent, characteristic of localization
of the vibrational modes (c), while the low frequency mode $\omega_L(q)$ increases
monotonically with increasing $q$, characteristic of collective dynamics (d).
The dashed line shown in panel (d) corresponds to the linear fit of the dispersions
at low-$q$ values.\label{kdyn}}
\end{figure}

The vibrational frequency $\omega$ as a function of $k$ is presented
in Fig.~\ref{kdyn}(a). Clearly, frequency $\omega$ is positively correlated
with $k$ at short time scale, that is, the larger the local connectivity, the
higher frequency the type of the particle motion. According to previous
study by Wakeda {\it et al.}~\cite{Wak-am}, the icosahedrally coordinated
particle with larger connectivity has a much higher average elastic rigidity,
thus giving a reasonable account for the positive trend showing in
Fig.~\ref{kdyn}(a) based on the well-known relationship between
vibrational frequency and elastic modulus.

The $k$-dependence of weighted parameters $C_{L/H}$, presented in
Fig.~\ref{kdyn}(b), show different behaviors as $k$ increases. The fraction
of the $\omega_L$ mode increase, while the fraction of the $\omega_H$
mode decreases. To identify whether the damped oscillation showing in
the ISF curve is a fully localized mode or acoustic-like mode with collective
behavior, we measure the incoherent ISF of overall icosahedrally coordinated
particles for values of $q$ in the range of 0.1 \r{A}$^{-1}$ $\le q\le$ 0.6 \r{A}$^{-1}$
and fit the ISF by the LCHO model. The obtained $\omega_H(q)$ and $\omega_L(q)$,
are shown in Fig.~\ref{kdyn}(c) and \ref{kdyn}(d), respectively.  It can be seen
that $\omega_H(q)$ is approximately $q$-independent in the entire investigated
$q$ range, suggesting that the high frequency mode is localized. In contrast,
$\omega_L(q)$ is markedly $q$-dependent and increases with increasing $q$,
indicating that the collective dynamics persist up to these frequencies.
As shown in Fig.~\ref{kdyn}(d), $\omega_L(q)$ displays linear behavior up
to $q$=0.35~\r{A}$^{-1}$. This implies that such excitations is acoustic phonon-like,
and the corresponding sound velocity is about 4082 m/s estimated based on
$v(q)=\omega_L(q)/q$, comparable with the experimental data reported for the
CuZr-based metallic alloys~\cite{Wang-pms}. Therefore, the short-time scale
behavior of ISF is a superposition of localized and extended mode with different
frequencies. That is, the damped oscillations showing at the short-time scale of
the ISF curves in our modeling system can be attributed to the dephasing of
localized and extended mode.

We point out that the well-defined acoustic mode exists up to 3 THz (Fig.~\ref{kdyn}(d)).
According to previous studies~\cite{Bunz-prl,Li-jmst}, the value of Boson peak frequency,
$\omega_{\rm BP}$, in CuZr alloys is in a frequency range of 2.5 to 3.5 THz. Therefore,
our results clearly show that the low frequency extended modes contribute to the
boson peak. Moreover, these results also show that the short-time valley in the
correlation functions and the vibrational boson peak dynamics are closely correlated.
\begin{figure}[ht]
\centering
\includegraphics[width=1.0\linewidth]{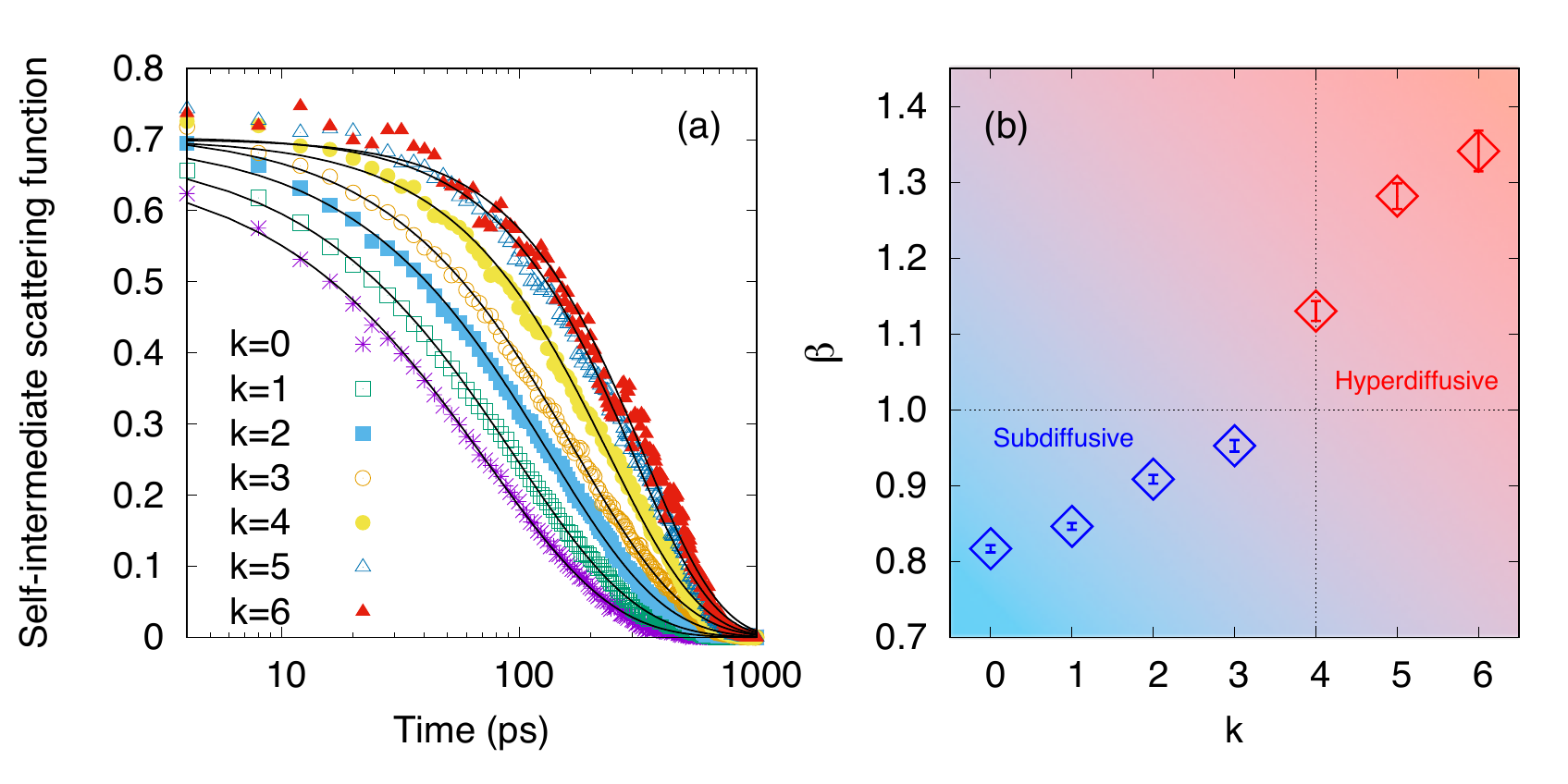}
\caption{Long-time behavior of the intermediate scattering function.
(a) The long-time decay of the correlation functions for particles with
different $k$ values. The black solid lines are the Kohlrausch-Williams-Watt (KWW) fits.
(b) The $k$ dependence of the exponent $\beta$. $\beta$ is large than 1 for
connectivity $k \ge $ 4 and less than 1 for $k<$ 4, revealing a dynamic
crossover from stretched exponential relaxation to
compressed one.\label{isfbeta}}
\end{figure}

We next focus on the behavior of the long-time structural relaxation for
particles with different values of $k$ by evaluating the corresponding ISFs.
The results are presented in Fig.~\ref{isfbeta}(a). The long-time decay of
the correlation function follows the Kohlrausch-Williams-Watt (KWW) expression,
$F_s(q,t)\propto A_q\exp[-(t/\tau)^{\beta}]$, where $A_q$ is the non-ergodicity
factor, $\tau$ is the relaxation time, and $\beta$ is the shape exponent.
Usually, the correlation function at long time scale can be well described by
a stretched exponential with exponent $\beta<$ 1 in supercooled liquids, a
signature of subdiffusive motion of the particles. However, as can be seen in
Fig.~\ref{isfbeta}(b), the shape of the ISF changes from stretched-exponential ($\beta<$ 1)
to compressed exponential ($\beta>$ 1) at $k=$ 4, which correspond to a dynamic
transition from subdisffusive motion to hyperdisffusive motion. Since $k=$ 4
enhance the construction of a $3$-dimensional network, this dynamic transition
has a geometric origin. Interestingly, the transition at $k=$ 4 here is similar to
the isostatic jamming transition with an average contact number of 4 in
$3$-dimensions for frictional spheres~\cite{Song-nat,Hecke-jpcm,Unger-prl}.

Compressed-exponential behavior in the correlation functions has been
widely reported as a common feature for out-of-equilibrium materials,
where the dynamics is controlled by hyperdiffusive motion. In glasses,
such hyperdiffusive motion is attributed to the release of internal stress
stored in the system during quenching~\cite{Cip-prl}, while for nanoparticles
suspended in a supercooled solvent, cooperative behavior governed by the
near-vitreous solvent is thought to be the origin of the faster-than-exponential
relaxations~\cite{Car-prl,Guo-prl}. In our study, cooperative motion and release
of internal stress coexist in our system, both of which may contribute to the
hyperdiffusive dynamics. Firstly, the developing of more cooperative motions
of particles is reflected by the positive correlation between the local connectivity
and the contribution from the extended mode shown in Fig.~\ref{kdyn}(b),
consistent with the picture of the hyperdiffusive behavior for suspensions of
nanoparticles~\cite{Car-prl,Guo-prl}. Secondly, according to previous
study~\cite{Wak-am}, isosahedrally coordinated particles have larger
average elastic modulus but smaller average atomic volume with increasing
local connectivity, thus particles with large connectivity ($k \ge $ 4) assume
more local stress (concentration of stress). Therefore, the hyperdiffusive
motion of the particles with large connectivity ($k \ge $ 4) in our study also
has contributions from the release of the local stress. This is similar to the
microscopic origin of the internal stress proposed for colloidal gels, where
the source of local stress was ascribed to the local deformation of the
elastic network due to the {\it syneresis} of the gel~\cite{Cip-prl}.

Previous studies demonstrated~\cite{Zhang-apl} that annealing at temperatures
slightly below the glass transition temperature effectively increases the fraction
of short-range icosahedral order in metallic glasses. Due to the self-aggregation
effect~\cite{Li-prb}, higher fraction of icosahedra is an indication of more particles
with local connectivity $k \ge $ 4, whose structural relaxation is a compressed
exponential decay. It is highly reminiscent of the very recent experimental
observation~\cite{Luo-prl} that pre-annealed metallic glasses at 0.9 $T_g$
show fast dynamic modes ($\beta>$ 1) exhibiting ballistic-like feature.
Researchers~\cite{Ruta-prl,Luo-prl} have attributed this faster-than-exponential
relaxation in metallic glasses to the presence of internal stresses induced
by quenching. Our results provide a possible hint of the structural origin
for the unexpected dynamics in glassy states.

To summarize, we provide a simulation evidence that metallic liquid
under supercooling exhibits quite rich dynamics. The particles `protected'
by particular spatial symmetries in fragile liquids possess dynamic
features that usually belong to strong glass formers. In addition,
the long-time relaxation of such particles decays as stretched or
compressed exponential which corresponds to either subdiffusive
or hyperdiffusive motion structurally depending on their local connectivity.
Therefore, the local connectivity is a good structural order parameter to
bridge multi-time scales dynamic processes in supercooled liquids and
glasses, which shows that the correlation between the fast vibrational
dynamics and the slow structural relaxation in glass-forming system
has an atomic-level structural origin. The coexistence of distinct dynamics
in one system sheds new light on understanding of the structure-dynamics
relationships in supercooled liquids and glasses.
%

\begin{acknowledgments}
We thank B. Ruta, M. Z. Li, P. F. Guan, and R. Blumenfeld for useful
discussion. This work is supported by the National Basic Research
Program of China (973 Program, Grant No. 2015CB856801),
National Natural Science Foundation of China (Grant No. 11525520),
and China Postdoctoral Science Foundation (Grant No. 2017M610687).
\end{acknowledgments}


\end{document}